\documentclass[preprint,superscriptaddress,showpacs,showkeys]{revtex4}

\usepackage{graphicx}
\usepackage{dcolumn}
\usepackage{bm}
\usepackage{setspace}

\def\ppbar{$p\overline{p}~$}             
\def\ttbar{$t\overline{t}~$}             

\def\pt{$p_T$}                          
\def\met{\mbox{${\hbox{$E$\kern-0.6em\lower-.1ex\hbox{/}}}_T$}} 
\def\mex{\mbox{${\hbox{$E$\kern-0.6em\lower-.1ex\hbox{/}}}_x$}} 
\def\mey{\mbox{${\hbox{$E$\kern-0.6em\lower-.1ex\hbox{/}}}_y$}} 
\def\mexy{\mbox{${\hbox{$E$\kern-0.6em\lower-.1ex\hbox{/}}}_{x,y}$}} 
 
\def\D0{D\O}                            
\def\etal{{\sl et al.}}                 

\newcommand{\pythia}{{\sc pythia}}

\newcommand{\alpgen}{{\sc alpgen}}
\newcommand{\geant}{{\sc geant}}

\newcommand{\madevent}{{\sc madevent}}

\newcommand{\evtgen}{{\sc EvtGen}}

\def\xsL{8.5}
\def\xsLStatU{0.6}

\def\xsLSystD{0.7}

\def\xs{\xsL}
\def\xsStatU{\xsLStatU}

\def\xsSystD{\xsLSystD}

\begin{document}

\title{
\Large\bf
Measurement of the \boldmath $t\overline{t}$ Production Cross Section
with an {\it in situ} Calibration of $b$-jet Identification Efficiency
}

\affiliation{Institute of Physics, Academia Sinica, Taipei, Taiwan 11529, Republic of China} 
\affiliation{Argonne National Laboratory, Argonne, Illinois 60439, USA} 
\affiliation{University of Athens, 157 71 Athens, Greece} 
\affiliation{Institut de Fisica d'Altes Energies, Universitat Autonoma de Barcelona, E-08193, Bellaterra (Barcelona), Spain} 
\affiliation{Baylor University, Waco, Texas 76798, USA} 
\affiliation{Istituto Nazionale di Fisica Nucleare Bologna, $^{bb}$University of Bologna, I-40127 Bologna, Italy} 
\affiliation{Brandeis University, Waltham, Massachusetts 02254, USA} 
\affiliation{University of California, Davis, Davis, California 95616, USA} 
\affiliation{University of California, Los Angeles, Los Angeles, California 90024, USA} 
\affiliation{Instituto de Fisica de Cantabria, CSIC-University of Cantabria, 39005 Santander, Spain} 
\affiliation{Carnegie Mellon University, Pittsburgh, Pennsylvania 15213, USA} 
\affiliation{Enrico Fermi Institute, University of Chicago, Chicago, Illinois 60637, USA}
\affiliation{Comenius University, 842 48 Bratislava, Slovakia; Institute of Experimental Physics, 040 01 Kosice, Slovakia} 
\affiliation{Joint Institute for Nuclear Research, RU-141980 Dubna, Russia} 
\affiliation{Duke University, Durham, North Carolina 27708, USA} 
\affiliation{Fermi National Accelerator Laboratory, Batavia, Illinois 60510, USA} 
\affiliation{University of Florida, Gainesville, Florida 32611, USA} 
\affiliation{Laboratori Nazionali di Frascati, Istituto Nazionale di Fisica Nucleare, I-00044 Frascati, Italy} 
\affiliation{University of Geneva, CH-1211 Geneva 4, Switzerland} 
\affiliation{Glasgow University, Glasgow G12 8QQ, United Kingdom} 
\affiliation{Harvard University, Cambridge, Massachusetts 02138, USA} 
\affiliation{Division of High Energy Physics, Department of Physics, University of Helsinki and Helsinki Institute of Physics, FIN-00014, Helsinki, Finland} 
\affiliation{University of Illinois, Urbana, Illinois 61801, USA} 
\affiliation{The Johns Hopkins University, Baltimore, Maryland 21218, USA} 
\affiliation{Institut f\"{u}r Experimentelle Kernphysik, Karlsruhe Institute of Technology, D-76131 Karlsruhe, Germany} 
\affiliation{Center for High Energy Physics: Kyungpook National University, Daegu 702-701, Korea; Seoul National University, Seoul 151-742, Korea; Sungkyunkwan University, Suwon 440-746, Korea; Korea Institute of Science and Technology Information, Daejeon 305-806, Korea; Chonnam National University, Gwangju 500-757, Korea; Chonbuk National University, Jeonju 561-756, Korea} 
\affiliation{Ernest Orlando Lawrence Berkeley National Laboratory, Berkeley, California 94720, USA} 
\affiliation{University of Liverpool, Liverpool L69 7ZE, United Kingdom} 
\affiliation{University College London, London WC1E 6BT, United Kingdom} 
\affiliation{Centro de Investigaciones Energeticas Medioambientales y Tecnologicas, E-28040 Madrid, Spain} 
\affiliation{Massachusetts Institute of Technology, Cambridge, Massachusetts 02139, USA} 
\affiliation{Institute of Particle Physics: McGill University, Montr\'{e}al, Qu\'{e}bec, Canada H3A~2T8; Simon Fraser University, Burnaby, British Columbia, Canada V5A~1S6; University of Toronto, Toronto, Ontario, Canada M5S~1A7; and TRIUMF, Vancouver, British Columbia, Canada V6T~2A3} 
\affiliation{University of Michigan, Ann Arbor, Michigan 48109, USA} 
\affiliation{Michigan State University, East Lansing, Michigan 48824, USA}
\affiliation{Institution for Theoretical and Experimental Physics, ITEP, Moscow 117259, Russia}
\affiliation{University of New Mexico, Albuquerque, New Mexico 87131, USA} 
\affiliation{Northwestern University, Evanston, Illinois 60208, USA} 
\affiliation{The Ohio State University, Columbus, Ohio 43210, USA} 
\affiliation{Okayama University, Okayama 700-8530, Japan} 
\affiliation{Osaka City University, Osaka 588, Japan} 
\affiliation{University of Oxford, Oxford OX1 3RH, United Kingdom} 
\affiliation{Istituto Nazionale di Fisica Nucleare, Sezione di Padova-Trento, $^{cc}$University of Padova, I-35131 Padova, Italy} 
\affiliation{LPNHE, Universite Pierre et Marie Curie/IN2P3-CNRS, UMR7585, Paris, F-75252 France} 
\affiliation{University of Pennsylvania, Philadelphia, Pennsylvania 19104, USA}
\affiliation{Istituto Nazionale di Fisica Nucleare Pisa, $^{dd}$University of Pisa, $^{ee}$University of Siena and $^{ff}$Scuola Normale Superiore, I-56127 Pisa, Italy} 
\affiliation{University of Pittsburgh, Pittsburgh, Pennsylvania 15260, USA} 
\affiliation{Purdue University, West Lafayette, Indiana 47907, USA} 
\affiliation{University of Rochester, Rochester, New York 14627, USA} 
\affiliation{The Rockefeller University, New York, New York 10065, USA} 
\affiliation{Istituto Nazionale di Fisica Nucleare, Sezione di Roma 1, $^{gg}$Sapienza Universit\`{a} di Roma, I-00185 Roma, Italy} 

\affiliation{Rutgers University, Piscataway, New Jersey 08855, USA} 
\affiliation{Texas A\&M University, College Station, Texas 77843, USA} 
\affiliation{Istituto Nazionale di Fisica Nucleare Trieste/Udine, I-34100 Trieste, $^{hh}$University of Trieste/Udine, I-33100 Udine, Italy} 
\affiliation{University of Tsukuba, Tsukuba, Ibaraki 305, Japan} 
\affiliation{Tufts University, Medford, Massachusetts 02155, USA} 
\affiliation{Waseda University, Tokyo 169, Japan} 
\affiliation{Wayne State University, Detroit, Michigan 48201, USA} 
\affiliation{University of Wisconsin, Madison, Wisconsin 53706, USA} 
\affiliation{Yale University, New Haven, Connecticut 06520, USA} 
\author{T.~Aaltonen}
\affiliation{Division of High Energy Physics, Department of Physics, University of Helsinki and Helsinki Institute of Physics, FIN-00014, Helsinki, Finland}
\author{B.~\'{A}lvarez~Gonz\'{a}lez$^v$}
\affiliation{Instituto de Fisica de Cantabria, CSIC-University of Cantabria, 39005 Santander, Spain}
\author{S.~Amerio}
\affiliation{Istituto Nazionale di Fisica Nucleare, Sezione di Padova-Trento, $^{cc}$University of Padova, I-35131 Padova, Italy} 

\author{D.~Amidei}
\affiliation{University of Michigan, Ann Arbor, Michigan 48109, USA}
\author{A.~Anastassov}
\affiliation{Northwestern University, Evanston, Illinois 60208, USA}
\author{A.~Annovi}
\affiliation{Laboratori Nazionali di Frascati, Istituto Nazionale di Fisica Nucleare, I-00044 Frascati, Italy}
\author{J.~Antos}
\affiliation{Comenius University, 842 48 Bratislava, Slovakia; Institute of Experimental Physics, 040 01 Kosice, Slovakia}
\author{G.~Apollinari}
\affiliation{Fermi National Accelerator Laboratory, Batavia, Illinois 60510, USA}
\author{J.A.~Appel}
\affiliation{Fermi National Accelerator Laboratory, Batavia, Illinois 60510, USA}
\author{A.~Apresyan}
\affiliation{Purdue University, West Lafayette, Indiana 47907, USA}
% ...
\collaboration{CDF Collaboration\footnote{With visitors from $^a$University of Massachusetts Amherst, Amherst, Massachusetts 01003,
$^b$Istituto Nazionale di Fisica Nucleare, Sezione di Cagliari, 09042 Monserrato (Cagliari), Italy,
$^c$University of California Irvine, Irvine, CA  92697, 
$^d$University of California Santa Barbara, Santa Barbara, CA 93106
$^e$University of California Santa Cruz, Santa Cruz, CA  95064,
$^f$CERN,CH-1211 Geneva, Switzerland,
$^g$Cornell University, Ithaca, NY  14853, 
$^h$University of Cyprus, Nicosia CY-1678, Cyprus, 
$^i$University College Dublin, Dublin 4, Ireland,
$^j$University of Fukui, Fukui City, Fukui Prefecture, Japan 910-0017,
$^k$Universidad Iberoamericana, Mexico D.F., Mexico,
$^l$Iowa State University, Ames, IA  50011,
$^m$University of Iowa, Iowa City, IA  52242,
$^n$Kinki University, Higashi-Osaka City, Japan 577-8502,
$^o$Kansas State University, Manhattan, KS 66506,
$^p$University of Manchester, Manchester M13 9PL, England,
$^q$Queen Mary, University of London, London, E1 4NS, England,
$^r$Muons, Inc., Batavia, IL 60510,
$^s$Nagasaki Institute of Applied Science, Nagasaki, Japan, 
$^t$National Research Nuclear University, Moscow, Russia,
$^u$University of Notre Dame, Notre Dame, IN 46556,
$^v$Universidad de Oviedo, E-33007 Oviedo, Spain, 
$^w$Texas Tech University, Lubbock, TX  79609, 
$^x$IFIC(CSIC-Universitat de Valencia), 56071 Valencia, Spain,
$^y$Universidad Tecnica Federico Santa Maria, 110v Valparaiso, Chile,
$^z$University of Virginia, Charlottesville, VA  22906,
$^{aa}$Yarmouk University, Irbid 211-63, Jordan,
$^{ii}$On leave from J.~Stefan Institute, Ljubljana, Slovenia, 
}}
\noaffiliation

\date{
June 25, 2010
}

\begin{abstract}
A measurement of the top-quark pair-production cross section
in \ppbar collisions at $\sqrt{s}=1.96$ TeV using data corresponding
to an integrated luminosity of 1.12 fb$^{-1}$
collected with the Collider Detector at Fermilab is presented.
Decays of top-quark pairs into the final states $e\nu$ +~jets and
$\mu\nu$ +~jets are selected, and the cross section and
the $b$-jet identification efficiency are determined using a new
measurement technique which requires
that the measured cross sections with exactly one
and multiple identified $b$-quarks from the top-quark decays agree.
Assuming a top-quark mass of 175 GeV/$c^2$, a cross section of
$\xs \pm \xsStatU (\rm {stat.}) \pm \xsSystD (\rm {syst.})~\rm{pb}$
is measured.
\end{abstract}



\maketitle

\section{Introduction}
\label{sec:introduction}
The properties of the top quark have been extensively studied since
its discovery by the CDF and D0 collaborations
at the Fermilab Tevatron in 1995~\cite{ref:cdftopdiscovery, ref:d0topdiscovery}. 
The \ttbar production cross section was measured in all detectable decay
channels, and good agreement was found between the results and the 
perturbative QCD calculations~\cite{ref:pdg}. 
At the Fermilab Tevatron, a \ppbar collider with a center-of-mass energy of 1.96 TeV, the
dominant standard model (SM) mechanisms for \ttbar production are $q\bar{q}$
annihilation ($85\%$) and gluon fusion ($15\%$). For a top-quark mass
of 175~GeV/$c^2$ the predicted total cross section is 
$\sigma_{t\overline{t}}=6.7_{-0.7}^{+0.6}$~pb~\cite{ref:cacciari}.
Previous measurements have been limited by statistical
uncertainties and by the uncertainty in the heavy-flavor jet identification
efficiency at high energy~\cite{ref:secvtx}. 

The top quark decays into a $W$ boson and a $b$ quark almost 100\%
of the time~\cite{ref:pdg}. The signal significance is expected to be greatest in the
lepton+jets channel, in which one $W$ decays leptonically and the other $W$
decays to quarks. These \ttbar events contain a high-momentum charged 
lepton, four jets from the four final-state quarks, and an undetected 
neutrino.
To enhance the \ttbar purity, at least one jet
in the event is usually required to be identified as originating from a bottom quark 
($b$-tagged)~\cite{ref:secvtx}.
The $b$-tagging efficiency, needed as an input to the measurement, introduces one 
of the largest sources of systematic uncertainty~\cite{ref:secvtx}.
In this Letter we present a new technique to measure the \ttbar cross section
(${\sigma_{t\overline{t}}}$) and determine the  $b$-tagging efficiency in lepton+jets events,
which reduces the systematic uncertainty of the measurement and allows a
determination of the $b$-tagging efficiency directly in the \ttbar sample.
This is the first use of the \ttbar sample for {\it in situ} calibration of the
$b$-tagging efficiency. The improvements in the associated systematic uncertainties
benefit directly other analyses, particularly searches for the Higgs boson~\cite{cdf_higgs} and 
supersymmetric particles~\cite{cdf_susy} or other scenarios of new physics~\cite{cdf_4th}.

\section{Selection of \ttbar candidate events}
\label{sec:selection}
Results reported here are obtained
using  1.12~fb$^{-1}$ of integrated luminosity collected
between March 2002 and August 2006 by
the Collider Detector at Fermilab (CDF II).
CDF II~\cite{ref:CDF_detector} is a general-purpose particle detector 
located at one of
the two interaction points of the Tevatron Collider.
Charged-particle tracking is provided by an eight-layer silicon
detector,
surrounded by a 3.1~m long open-cell drift chamber,
the central outer tracker (COT).
Both are contained in a superconducting solenoid with a 1.4~T magnetic field.
The silicon system provides three-dimensional
hit information between radii of 1.4~cm and 28~cm, 
and allows to measure the distance of closest approach of energetic
tracks to the event vertex in the transverse plane (impact parameter, $d_0$)
with a resolution of $\sim$ 40~$\mu$m, including a 30~$\mu$m\
contribution from the beamspot.
The excellent impact parameter resolution is critical to identify
displaced tracks which are associated to bottom quarks resulting from 
top-quark decays.
The COT covers the pseudorapidity~\cite{ref:coord} range $|\eta| < 1.1$ and
provides a long lever arm for track curvature measurements.
Outside the solenoid, electromagnetic and hadronic
calorimeters arranged in projective towers surround the tracking volume and 
absorb photons, electrons and hadrons with $|\eta| < 3.6$.
Beyond the calorimeters, drift chambers track penetrating muons in the region 
$|\eta| < 1.0$.

The data were collected with two high-$p_T$ lepton triggers,
one of which requires a high-$E_T$ electron ($E_T > 18$ GeV) 
and the other a high-$p_T$ muon ($p_T > 18$ GeV/$c$).
The trigger efficiency is 95.3 $\pm$ 1.5\% (89.1 $\pm$ 1.6\%) for identified 
electrons (muons).
Events are 
selected off-line by requiring the presence of an
isolated~\cite{ref:iso} electron (muon) candidate with
$E_T > 20$~GeV and $|\eta| < 1.1$ ($p_T > 20$~GeV/$c$ and $|\eta| < 1.0$), and 
at least three jets with $E_T > 20$ GeV and $|\eta|<2$.
Jets are clustered with a cone-based algorithm with a 
cone size
$\Delta R \equiv \sqrt{\Delta \phi^2 + \Delta \eta^2} =0.4$,
and their energies are corrected for instrumental effects
and excess energy from additional \ppbar collisions~\cite{ref:jes}.
The primary vertex position along the beam is required to lie inside the
luminous region ($|z|<60$~cm) and to be consistent with the $z$ position of 
the point of origin of the high-\pt\ lepton. 
Events with additional identified  high-\pt\ leptons, or a track 
which forms a value compatible with the $Z$ boson mass if combined with 
the primary lepton, are removed to suppress backgrounds.
To account for the expected neutrino, we require large missing transverse energy~\cite{ref:coord},
$\met > 20$~GeV, which rejects $\sim 50\%$ of background events that do not contain
a real $W$ boson. 
Finally, as \ttbar events typically have larger total transverse energy than
background events, we require the $H_T$~\cite{ref:coord} to exceed 200~GeV.
Table~\ref{tab:singletag} includes the event count before $b$-tagging (pretag) sorted 
by the number of jets in the event.
In the one- and two-jet bins, where background dominates
over the signal, the $H_T$ requirement is not applied
and the samples are used as control samples.

\section{Identification of $b$-jets}
\label{sec:btagging}
Requiring at least one jet to be $b$-tagged considerably reduces the
background.  The 
$b$-tagging algorithm exploits the long lifetime of
bottom hadrons by identifying decay vertices inside jets~\cite{ref:secvtx}.
These vertices are reconstructed requiring a minimum of two or three
tracks with an impact parameter significance ($d_{0}/\sigma_{d_{0}}$) greater than 3.0
or 2.0, respectively.  Track combinations consistent with a $K^0_S$ or
$\Lambda$ are removed, and an upper limit of $d_0 = 0.15$~cm is used to
reject interactions with detector material.  We measure the two-dimensional
displacement of the secondary vertex from the primary interaction point
projected along the jet axis ($L_{2D}$).  A jet is $b$-tagged if the vertex
has $L_{2D}$ significance ($L_{2D}/\sigma_{L_{2D}}$) larger 
than 6.0, where the uncertainty on $L_{2D}$
includes contributions from both the primary and secondary vertex fits.
The probability of misidentifying a light-flavor jet as a $b$-quark jet
due to detector resolution (mistag rate) is $1.9\pm0.4\%$, estimated from
secondary vertices reconstructed on the opposite side of the primary vertex with $L_{2D}$
significance less than $-6.0$ in a generic jet sample~\cite{ref:thesis_daniel}. 
The mistag rate is corrected by a factor of $1.3\pm0.1$ to account for
the remaining contribution of long-lived
light-flavor hadrons ($K_S$ and $\Lambda$) and material interactions
that are present only at positive $L_{2D}$.

The \ttbar acceptance calculation is based on the \pythia~\cite{ref:pythia}
Monte Carlo simulation, with the CTEQ5L parton distribution
functions~\cite{ref:cteq5l} and assuming a top mass of 175~GeV/$c^2$. Heavy-flavor 
decays are treated by \evtgen~\cite{ref:evtgen}. Monte Carlo events
are passed through a \geant~\cite{ref:geant} simulation of the detector
and subjected to the same selection requirements as the data.  
The total acceptance before $b$-tagging is $4.3\pm0.1\%$ ($3.5\pm0.1\%$)
for electron (muon) events and includes the branching fraction
of the $W$ boson, the geometric and kinematic acceptances, and the lepton
identification and trigger efficiencies.

In previous cross section measurements~\cite{ref:secvtx} the 
tagging efficiency was 
needed as an input in order to perform the measurement. In those cases, the 
efficiency for the full \ttbar event was determined from simulation,
and a multiplicative scale factor ($S_b$) was applied to the efficiency found
in the simulation to correct for the per-jet efficiency
difference between data and simulation. This difference is caused by
imperfections in the simulation, arising from, for example, incomplete 
description of the silicon detector, tracking efficiencies, and hadron decay modeling. 
The scale factor $S_b$ was measured using high statistics data and simulation
samples, enriched in heavy flavor by requesting non-isolated low-$p_T$ leptons,
and used to correct the heavy-flavor tagging efficiency in all the simulated
samples.
In the current 1.12~fb$^{-1}$ data sample, $S_b$ has been measured to be
$S_b = 0.95 \pm 0.05$, where the uncertainty is dominated by
the extrapolation from the low-$p_T$ calibration sample
to typical \ttbar jet energies~\cite{ref:thesis_daniel}. 
Using this nominal scale factor in a \ttbar sample we 
expect $46\pm4\%$ of the events to have 
exactly one $b$-tagged jet and $22\pm3\%$ to have
two or more $b$-tagged jets.
Charm-quark jets which are $b$-tagged are treated analogously to 
$b$-quark jets, but we assume a 10\% uncertainty on the scale factor for charm, $S_c$.

\section{Background Estimation}
\label{sec:background}
The primary background process is direct $W$ production with multiple jets,
where the $W$ boson decays leptonically. Smaller contributions come from
QCD jet production in which the $W$ signature is faked by jets appearing as electrons
or by semi-leptonic $b$-hadron decays (non-$W$), and different electroweak processes:
single-top-quark production, diboson ($WW$, $WZ$, and $ZZ$) production,
and $Z$ boson decays to tau pairs.

The $W$+jets contribution to the background is separated into events
with and without heavy-flavor jets. 
To estimate the background contribution from $W$ events with only
light-flavor jets ($N^{W+LF}_{\rm tag}$), the mistag rate is parametrized as a function
of jet $E_T$,  $\eta$, and number of tracks, as well as event total energy, number of 
primary vertices, and primary vertex $z$ position.
This parametrization is applied to the pretag
dataset, and the
result is scaled down to the fraction of the data not attributed 
to a physics process with heavy-flavor.

To determine the number of tags that originated from $W$ events
accompanied by heavy-flavor quarks,
the fractions of $W$+jets events attributable to $Wb\bar{b}$, $Wc\bar{c}$, and $Wc$
are estimated with \alpgen\ Monte Carlo~\cite{ref:alpgen}.
Since in the simulation heavy-flavor production may arise both from the
soft-radiation evolution of a given event or directly from
the parton configuration, a procedure to avoid double counting
has been implemented~\cite{ref:thesis_daniel}.
The fractions of $W$+jets with heavy flavor (heavy-flavor fractions, $F_i$, where
the index corresponds to the different contributions)
are calibrated using data and simulated samples of inclusive jets.
The bottom and charm fractions in the simulation are extracted from 
the generator information, while the equivalents in the data are 
obtained from template fits to the kinematic and dynamic properties
of the  
tagged events. 
The measured heavy-flavor fraction calibration
factor is consistent with unity, except in the one-jet bin, where
the value for the data is about $30\%$ higher, the difference is assigned
as an uncertainty~\cite{ref:thesis_daniel}. 
The expected number of $W$ events with heavy-flavor ($N^{W+HF}_{\rm{tag}}$) is 
estimated by multiplying these fractions by the number of pretag events
($N^{W}_{\rm{pre}}$)
and the tagging efficiency in these events ($\epsilon_i$), measured from 
simulation and corrected by the scale factor, 
$N^{W+HF}_{\rm{tag}}=N^{W}_{\rm{pre}} \sum_i \epsilon_i F_i$.
The contribution of other backgrounds and the $t\bar{t}$ signal
are removed from the $N^{W}_{\rm{pre}}$ pretag expectation:
$N^{W}_{\rm{pre}}=N^{\rm{data}}_{\rm{pre}}-N^{non-W}_{\rm{pre}}-
N^{t\bar{t}}_{\rm{pre}}-N^{\rm{ewk}}_{\rm{pre}}$,
where $N^{\rm{data}}_{\rm{pre}}$ is the number of pretag events,  $N^{t\bar{t}}_{\rm{pre}}$ is 
the number of pretag events attributed to \ttbar production,
$N^{non-W}_{\rm{pre}}$ is the number of events without a $W$ boson, and
$N^{\rm{ewk}}_{\rm{pre}}$ is the contribution from single-top-quark production, $Z$ boson
decays to tau pairs, and diboson production. Both $N^{non-W}_{\rm{pre}}$ and
$N^{\rm{ewk}}_{\rm{pre}}$ are discussed below.

Some events without a $W$ boson may satisfy the pretag requirements of an 
identified lepton and large missing energy. For example, photon conversions or 
mis-identified pions or kaons may comply with the lepton requirements, and 
the missing energy may arise due to mis-measured jets. 
Heavy-flavor di-jet production may also result in leptons from the semi-leptonic
decay faking the $W$ boson signal. 
The expectation for this non-$W$ background is determined from data, by fitting 
the missing energy distribution in each jet multiplicity bin.
The templates used in the fit are derived from \pythia\ 
($t\overline{t}$) and \alpgen\ ($W$+jets) Monte Carlo,
and samples enriched in fake leptons and heavy flavor. 
The sample with fake leptons is obtained from events in data 
where the primary lepton fails at least two 
identification requirements~\cite{anti-electron}, while
di-jet events with non-isolated low-$E_T$ leptons 
provide the heavy-flavor enriched sample.
The \ttbar contribution is fixed to the value expected
for the measured cross section. The fraction of non-$W$ events
is measured from the fits both before and after requiring a $b$-tagged jet.

A Monte Carlo-based method is used to estimate the remaining backgrounds due
to single top quark production (\pythia/\madevent~\cite{ref:madevent}), vector boson pair production (\pythia)
and $Z\rightarrow\tau\tau$ (\alpgen\ interfaced to \pythia), normalizing the
expectations to their respective theoretical cross sections~\cite{ref:theoxsec}.
The tagging efficiency
is taken from these Monte Carlo samples and corrected by 
$S_b$.

\section{Results}
\label{sec:results}
\begin{figure}[htb]
\begin{center}
\includegraphics[width=2.9in]{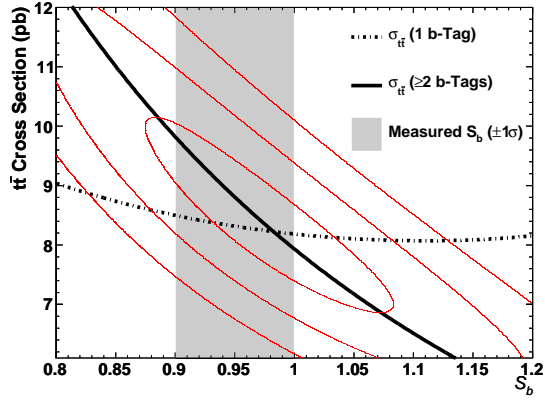}
\caption{Dependence of the exclusive single and multiple tagged cross sections on the
$b$-tagging efficiency scale factor. The point where the two measurements intersect is 
in good agreement with the 
independent determination of the $b$-tag scale factor from low-$p_T$ data. The associated
confidence bands are also shown.  
}
\label{fig:sfxs_fit}
\end{center}
\end{figure}

\begin{center}
\begin{figure}[htb]
\includegraphics[width=3.1in]{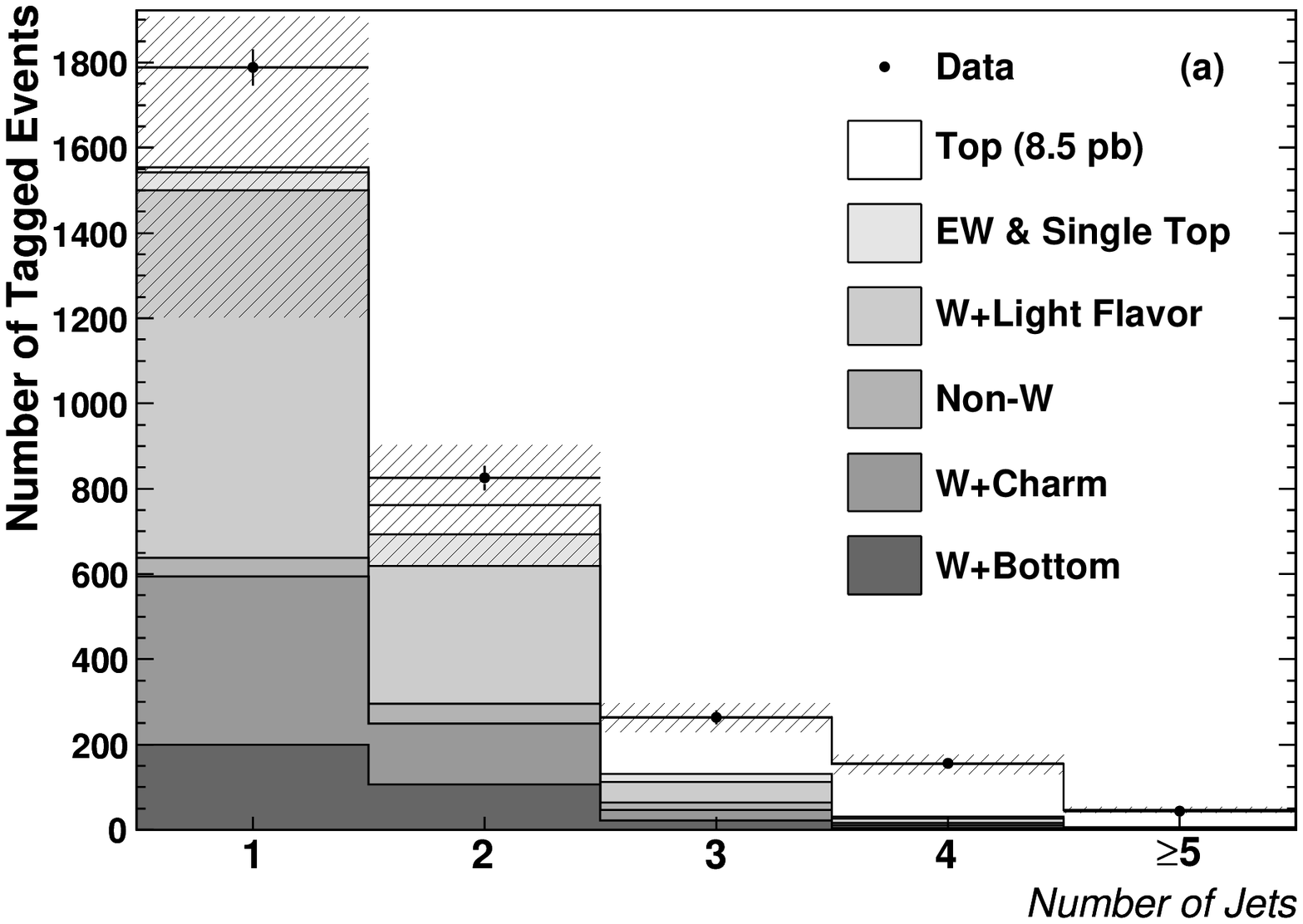}
\includegraphics[width=3.1in]{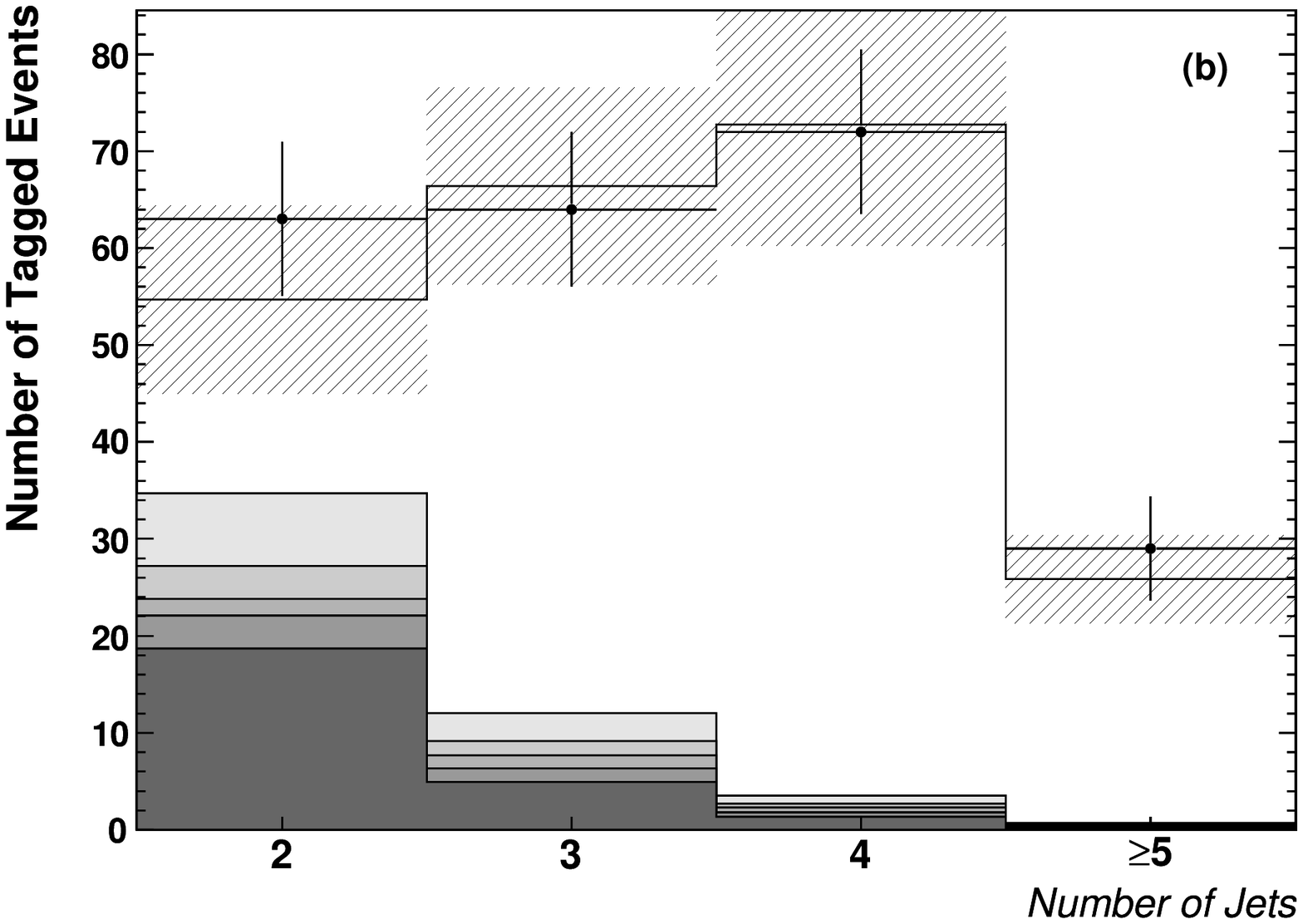}
    \caption{Summary of background and signal event yields versus number of
   jets in the event when requiring (a) one $b$-tagged jet
   and (b) at least two $b$-tagged jets.
   The \ttbar contribution is normalized to the measured cross section.
   The $H_T$ requirement is released for events with fewer than 3~jets.
   The hashed region shows the total systematic uncertainty on the expectation.}
    \label{fig:composition}
\end{figure}
\end{center}

The scale factor enters the equation in the signal term as well as in 
the $W$+jets with heavy-flavor and electroweak backgrounds. 
Instead of using the $S_b$ value measured in the low-$p_T$
lepton sample, in this paper we perform a simultaneous fit
of the \ttbar cross section and $S_b$ by requiring that the single and
multiple-tagged samples correspond to the same \ttbar cross section.
Since the $W$+jets backgrounds and, to a lesser extent, the non-$W$
background depend on the assumed \ttbar cross section, the procedure to 
determine them is iterative. 
The \ttbar contribution to the $b$-tagged sample is written as
$N^{t\bar{t}}_{\rm{tag}}=N_{\rm{tag}}-N_{\rm{tag}}^{W+HF}-N_{\rm{tag}}^{W+LF}-N_{\rm{tag}}^{\rm{ewk}}$.
Figure~\ref{fig:sfxs_fit} shows the
cross section measurement for exclusive single $b$-tagged events,
and for events with two or more $b$-tagged jets
as a function of $S_b$. 
A Poisson likelihood fit yields results of $\sigma_{t\bar{t}}=8.2\pm0.9$ pb 
with only the statistical and scale factor uncertainties included, and 
$S_b=0.98\pm0.07$, consistent with the $S_b$ value 
measured in the low-$p_T$ lepton sample.
The associated confidence bands are also shown in Fig.~\ref{fig:sfxs_fit}.
The signal and backgrounds
contributions to the $b$-tagged data sample are summarized in 
Tables~\ref{tab:singletag} and~\ref{tab:doubletag} for the single and the multiple-tagged
data respectively, and are shown in Fig.~\ref{fig:composition} as a function of the jet 
multiplicity.

The systematic uncertainties on the cross section, before $b$-tagging
requirements, are dominated by a $4.2\%$ uncertainty due to the jet energy
calibration.  Other sources of uncertainty are 
the choice of Monte Carlo generator ($2.1\%$),
the lepton identification efficiency ($2.0\%$),
the choice of parton distribution functions ($0.7\%$),
and the modeling of initial and final state radiation ($0.5\%$).
The uncertainty in the integrated luminosity
is $6\%$~\cite{ref:lum}. A $3.3\%$ uncertainty associated to the background estimation
was determined using a large ensemble of simulated experiments by fluctuating 
the backgrounds within their uncertainties. The different sources of systematic 
uncertainty are added in quadrature for the signal expectation~\cite{ref:secvtx}.

In order to include the 
independent measurement of
the $S_b$ in the lepton sample ($S_b=0.95\pm0.05$), a term to penalize 
deviations from this value is added to the likelihood. With this
constraint we measure a cross section of 
$\xs \pm \xsStatU (\rm {stat.}) \pm \xsSystD (\rm {syst.})~\rm{pb}$
and a scale factor of $0.96\pm 0.04$.  With this $S_b$ 
the fraction of selected \ttbar candidates with exactly one
identified $b$-quark is $46\pm4\%$ and with two or more $b$-quarks
is $23\pm3\%$.  
The statistical uncertainty of $\sigma_{t\bar{t}}$ also includes 
the uncertainty on $S_b$.

In conclusion, we have performed the first simultaneous fit
of the \ttbar production cross section and the $b$-tagging
efficiency in \ppbar collisions at $\sqrt{s}=1.96$ TeV
using data corresponding to an integrated luminosity of 1.12~fb$^{-1}$. 
The cross section result, 
$\xs \pm \xsStatU (\rm {stat.}) \pm \xsSystD (\rm {syst.})~\rm{pb}$,
is consistent with the SM expectation of 
$6.7_{-0.7}^{+0.6}$~pb for a mass of 175~GeV/$c^2$. 
The dependence of the acceptance on
the top mass results in a variation of the measured cross section by 
$\pm0.1$ pb for every $\mp1$ GeV/$c^2$ shift in the assumed top-quark mass.
With the innovative technique presented in this Letter 
the $b$-tagging efficiency for high-$E_T$ $b$-jets was
directly measured in the \ttbar sample, and its uncertainty
reduced 
with respect to previous results~\cite{ref:secvtx}. 
Future measurements at the Large Hadron Collider, like searches 
for the Higgs boson and supersymmetric particles, where $b$-tagging
performance is critical, could benefit from this technique.

\section*{Acknowledgements}
We thank the Fermilab staff and the technical staffs of the participating
institutions for their vital contributions. This work was supported by
the U.S. Department of Energy and National Science Foundation; the Italian
Istituto Nazionale di Fisica Nucleare; the Ministry of Education, Culture,
Sports, Science and Technology of Japan; the Natural Sciences and Engineering
Research Council of Canada; the National Science Council of the Republic of
China; the Swiss National Science Foundation; the A.P. Sloan Foundation;
the Bundesministerium f\"ur Bildung und Forschung, Germany; the World
Class University Program, the National Research Foundation of Korea; the
Science and Technology Facilities Council and the Royal Society, UK; the
Institut National de Physique Nucleaire et Physique des Particules/CNRS;
the Russian Foundation for Basic Research; the Ministerio de Ciencia e
Innovaci\'{o}n, and Programa Consolider-Ingenio 2010, Spain; the Slovak
R\&D Agency; and the Academy of Finland.

\begingroup
\begin{table*}[hbt]
\begin{center}
\footnotesize 
\begin{tabular}{lccccc}\hline\hline
   & $W + 1$ jet  & $W + 2$ jets & $W + 3$ jets & $W + 4$ jets & $W + \geq$ 5 jets\\
\hline
 Pretag     & 78,903           &         12,873   &     1515           &         507    &    132 \\  \hline
Electroweak& 43  $\pm$ 5     &  75 $\pm$ 10    &   18 $\pm$ 2       & 4.9 $\pm$ 0.6  & 1.1 $\pm$ 0.1\\
 $W$+HF     & 594 $\pm$ 237   &  249 $\pm$ 100  &   47 $\pm$ 19      & 11 $\pm$ 4     & 1.7 $\pm$ 0.7\\
 $W$+LF     & 862 $\pm$ 259   &  323 $\pm$ 97   &   49 $\pm$ 15      & 9 $\pm$ 3      & 1.7 $\pm$ 0.5\\
 Non-$W$    & 44  $\pm$ 17    &  46 $\pm$ 19    &   17 $\pm$ 7       & 6 $\pm$ 2      & 1.5 $\pm$ 0.6\\  \hline
 Background  & 1542 $\pm$ 352  &  693 $\pm$ 141  &       131 $\pm$ 25 & 31 $\pm$ 6     & 6 $\pm$ 1\\
 $t\bar{t}$ & 11  $\pm$ 2     &  68 $\pm$ 12    &    132 $\pm$ 13    & 123 $\pm$ 12   & 40 $\pm$ 4\\ \hline
 Total      & 1554  $\pm$ 352 &  761 $\pm$ 141  &       263 $\pm$ 28 & 154 $\pm$ 13   & 46 $\pm$ 4\\
 Data events& 1788            & 825             &         264        & 156            &43 \\ \hline\hline
\end{tabular}
\caption{
Summary of event yields and background (Bkgd) expectations
sorted by the number of jets in the event.
Event totals before $b$-tagging (Pretag) are listed in the first row;
all other entries correspond to the sample with exactly one $b$-tagged jet,
assuming the measured \ttbar cross section of 8.5~pb.
For events with more than 2 jets the $H_T>200$~GeV requirement is applied. }
\label{tab:singletag}
\end{center}
\end{table*}
\endgroup

\begingroup
\begin{table}[htb]
\begin{center}
\begin{tabular}{lcccc} \hline\hline
            & $W + 2$ jets     & $W + 3$ jets      & $W + 4$ jets       & $W + \geq$ 5 jets\\ \hline
Electroweak& 7 $\pm$ 1      &  2.9 $\pm$ 0.4&  0.9 $\pm$ 0.1 & 0.22 $\pm$ 0.03 \\
 $W$+HF     & 22 $\pm$ 9     &  6 $\pm$ 2    &  1.8 $\pm$ 0.7 & 0.3 $\pm$ 0.1 \\
 $W$+LF     & 3 $\pm$ 1      &  1.4 $\pm$ 0.4&  0.4 $\pm$ 0.1 & 0.09 $\pm$ 0.03 \\
 Non-$W$    & 1.7 $\pm$ 0.7  &  1.4 $\pm$ 0.5&  0.5 $\pm$ 0.2 & 0.12 $\pm$ 0.05 \\ \hline
 Background & 35 $\pm$ 9     &  12 $\pm$ 3   &  3.5 $\pm$ 0.8 & 0.7 $\pm$ 0.1 \\
 $t\bar{t}$ & 20 $\pm$ 4     &  54 $\pm$ 10  &  69 $\pm$ 12   & 25 $\pm$ 4 \\ \hline
 Total      & 55 $\pm$ 10    &  66 $\pm$ 10  &  73 $\pm$ 12   & 26 $\pm$ 4 \\
 Data events& 63             &  64           &  72            & 29 \\  \hline\hline
\end{tabular}
\caption{
Summary of event yields and background (Bkgd) expectations
sorted by the number of jets in the event,
for events with at least  two $b$-tagged jets,
assuming the measured \ttbar cross section of 8.5~pb.
For events with more than 2 jets the $H_T>200$~GeV requirement is applied. 
}
\label{tab:doubletag}
\end{center}
\end{table}
\endgroup

\begin{thebibliography}{99}
\bibitem{ref:coord}
A cylindrical coordinate system with the $z$ axis along the proton 
direction is used, in which $\theta$ is the polar angle. We define
$E_T=E\,\sin\theta$, $p_T=p\,\sin\theta$, and pseudorapidity  
$\eta= -\ln \tan(\theta/2)$. 
The missing transverse energy is defined by 
$\met=-\sum_i E_T^i \hat{n}_i$, where $i$ is the calorimeter tower number and
$\hat{n}_i$  is a unit vector perpendicular to the beam axis and pointing
at the $i^{th}$ calorimeter tower. The missing transverse energy is
corrected for jets and muons. The $H_T$ is defined by the scalar sum of the $\met$, 
the lepton transverse momentum, and the total jet transverse energy.
\bibitem{ref:iso}
The energy inside
a cone of radius $R=\sqrt{(\Delta\phi)^2+(\Delta\eta)^2}=0.4$ around the
lepton is required to be less than $10\%$ of the electron (or muon) $E_T$ ($p_T$).
\end{thebibliography}
\end{document}